\documentclass[final,3.5p,onecolumn,times]{elsarticle}

\usepackage{amssymb,amsmath,psfrag,epsfig}
\usepackage{cancel}

\newcommand{\bea}{\begin{eqnarray}}
\newcommand{\eea}{\end{eqnarray}}
\newcommand{\bean}{\begin{eqnarray*}}
\newcommand{\eean}{\end{eqnarray*}}

\def\O #1{\overline{#1}}

\def\W #1{\widetilde{#1}}

\def\braket#1{\left\langle #1 \right\rangle}

\def\ket#1{\left| #1\right\rangle}

\def\b{{\beta}}

\def\vev{\braket}

\def\bvev#1{\left[ #1 \right]}
\def\Spaa{\vev}
\def\Spbb{\bvev}

\def\Label#1{\label{#1}%
  \smash{\hbox to0pt{\raise1ex\hbox{\tiny[#1]}\hss}}}

\begin{document}
\title{KLT and New Relations for $\mathcal{N}=8$ SUGRA and $\mathcal{N}=4$ SYM}

\author{{Bo Feng$^{a,c}$}, {Song He$^{b,c}$}\bigskip}

\address{$^a$\small Center of Mathematical Science, Zhejiang
University, Hangzhou, China \\
$^b$\small Max-Planck-Institut f\"{u}r Gravitationsphysik,
Albert-Einstein-Institut, Golm, Germany\\
$^c$\small Kavli Institute for Theoretical Physics China, CAS,
Beijing 100190, China \\}

\begin{abstract}
In this short note, we prove the supersymmetric Kawai-Lewellen-Tye
(KLT) relations between ${\cal N}=8$ supergravity (SUGRA) and ${\cal
N}=4$ super Yang-Mills (SYM) tree-level amplitudes in the frame of
S-matrix program, especially we do not use string theory or the
explicit Lagrangian form of corresponding theories. Our
supersymmetric KLT relations naturally unify the non-supersymmetric
KLT relations and newly discovered gauge theory identities and
produce more identities for amplitudes involving scalars and
fermions. We point out also that these newly discovered identities
can be used to reduce helicity basis from $(n-3)!$ further down.

\end{abstract}

\maketitle

\section{Introduction}

S-matrix program~\cite{S-matrix} is a program to study properties of
quantum field theory based on some general principles, like the
Lorentz invariance, Locality, Causality, Gauge symmetry as well as
Analytic property. Because it does not use specific information like
Lagrangian, result obtained by this method is quite general. Also
exactly because its generality with very few assumptions, study
along this line is very challenging.

There are three important progresses in the frame of S-matrix
program worth to mention. The first one is the unitarity cut method
proposed in~\cite{Bern:1994zx}, where on-shell tree amplitudes have
been used for calculations of loop amplitudes without drawing many
many Feynman diagrams. The second one is the BCFW on-shell recursion
relations~\cite{Britto:2004ap,Britto:2005fq}. The derivation of BCFW
relations beautifully demonstrates the idea of S-matrix program. It
relies only on basic analytic properties of the tree-level
amplitude: the single pole structure and the factorization property
when a propagator reaches its mass-shell. Then using the familiar
complex analysis the whole amplitude can be uniquely fixed if there
are no boundary contributions\footnote{The boundary behavior is one
important subject to study. In~\cite{ArkaniHamed:2008yf}, background
field method has been applied to the study. In~\cite{boundary}, the
situation with nonzero boundary contributions has also been
discussed. It will be interesting to study the boundary behavior in
the frame of S-matrix program.}.

Based on the BCFW recursion relations, the third beautiful work
along the line of S-matrix program is given by Benincasa and Cachazo
~\cite{Paolo:2007}. In the paper, by assuming the applicability of
BCFW recursion relations for four-particle amplitudes, they have
easily re-derived many well-known (but difficult to prove)
fundamental facts in S-matrix program, such as the non-Abelian
structure for gauge theory and all matters couple to gravity with
same coupling constant. Given the non-Abelian structure, the
applicability of BCFW recursion relations has been proved, using
purely S-matrix argument, for gauge theory amplitudes with arbitrary
number of particles~\cite{SHST}.

One important observation in~\cite{Paolo:2007} is that three point
amplitude (on-shell) of any theory is uniquely fixed by the Lorentz
invariance and the spin of external particles. For example, the
three gluon amplitudes are given by
\bea A_3(1^-,2^-,3^+)= {\Spaa{1|2}^3\over \Spaa{2|3}\Spaa{3|1}},~~~
A_3(1^+,2^+,3^-)= {\Spbb{1|2}^3\over
\Spbb{2|3}\Spbb{3|1}}~,~~~~\label{A3}\eea
while the three graviton amplitude is given by
\bea M_3(1,2,3)= A_3(1,2,3)^2~.~~~~\label{M3}\eea
Another result determined by Lorentz invariance and spin symmetry is
\bea A_3(1,2,3^+) A_3(1,2,3^-)=0~.~~~~\label{A3-vanish} \eea
It is worth to emphasize that the vanishing of (\ref{A3-vanish}) is
because for three massless particles, on-shell condition requires
either $\Spaa{1|2}\sim\Spaa{2|3}\sim \Spaa{3|1}= 0$ or
$\Spbb{1|2}\sim\Spbb{2|3}\sim \Spbb{3|1}= 0$. For three gluons, no
matter which helicity configuration it is, we will always have
$\sum_{i=1,2,3} h_i\neq 0$, thus amplitude will contain either
$\Spaa{~|~}$ or $\Spbb{~|~}$ depending on the sum of helicities to
be negative or positive. If the sum of helicities is zero
$\sum_{i=1,2,3} h_i= 0$, the situation is very tricky and systematic
exploration of this particular case is still missing.

The idea of~\cite{Paolo:2007} has intrigued several important works
in last few months. Staring form the antisymmetry of (\ref{A3}) plus
the validity of BCFW recursion relations, four important properties
of color-ordered gluon amplitudes have been proved in the frame of
S-matrix program in~\cite{Feng:2010my}. They are color-ordered
reversed relations, $U(1)$-decoupling relations, Kleiss-Kuijf
relations~\cite{Kleiss:1988ne} and Bern-Carrasco-Johansson(BCJ)
relations ~\cite{Bern:2008qj}\footnote{The BCJ relations have also
been proved in string
theory~\cite{BjerrumBohr:2009rd,Stieberger:2009hq}.}. In their
proof, there are no need for those inputs, such as Lagrangian
definition and string theory. In other words, it is possible to have
a deformed Yang-Mills Lagrangian and same results will hold  as long
as the BCFW recursion relations can be applied\footnote{The BCJ
relations have also been generalized to include the matter and to
the ${\cal N}=4$ supersymmetric theory in
~\cite{Sondergaard:2009za,Jia:2010nz} along the same line.}.

Using similar ideas, based only on observations (\ref{M3}) and
(\ref{A3-vanish}), as well as the validity of BCFW recursion
relations for graviton amplitudes
\cite{Benincasa:2007qj,ArkaniHamed:2008yf}, new form of
Kawai-Lewellen-Tye type relations~\cite{Kawai:1985xq} and new gauge
amplitude identities have been found and proved in
~\cite{BjerrumBohr:2010ta,BjerrumBohr:2010zb}. Again, in the proof,
it is no need for input, such as Einstein Lagrangian or string
theory. The BCJ relations are used in the proof, but since BCJ
relations have been proved in the frame of S-matrix, results
obtained in~\cite{BjerrumBohr:2010ta,BjerrumBohr:2010zb} are nicely
fit in the S-matrix program.

In this short note, we will generalize the result in
~\cite{BjerrumBohr:2010ta,BjerrumBohr:2010zb} to the case of ${\cal
N}=8$ SUGRA. Especially we will show that the KLT relations and the
new gauge theory identities can be unified into the ${\cal N}=8$ KLT
relations. With this unified form we can get even more identities
involving scalars and  fermions. We want to emphasize that although
the ${\cal N}=8$ KLT relations have been discussed in
~\cite{Bianchi:2008pu}, our study in this note is in the frame of
S-matrix program, i.e., we do not use the string theory to derive
the ${\cal N}=8$ KLT relations. Instead, we start from the three
point amplitude of ${\cal N}=8$ SUGRA and use BCFW recursion
relations to derive and prove general ${\cal N}=8$ KLT relations
purely from the point of view of field theory.

The plan of the note is the following. In section two, we write down
and prove the supersymmetric KLT relations between ${\cal N}=8$
SUGRA and ${\cal N}=4$ SYM. In section three, we discuss how the new
supersymmetric KLT relations can be used to produce many new
identities for helicity amplitudes. In section four, we give a brief
summary and some discussions.

\section{BCFW proof of KLT relations for ${\cal N}=8$ SUGRA}

One basic property of supersymmetric field theory is that different
fields are grouped into a supermultiplet. With such grouping, the
type and the helicity of fields are represented by the expansion of
a superfield in terms of supersymmetry Grassmann variables $\eta^A$
where $A=1,..,{\cal N}$ and ${\cal N}$ is the number of total
supersymmetries, thus there is $SU(\mathcal{N})$ $R$-symmetry with
$2^{\mathcal{N}}$ on-shell states. For example, the on-shell ${\cal
N}=4$ superfield is given by ~\cite{ArkaniHamed:2008gz,Drummond}
\begin{equation}\label{multiplet}
\Phi(p,\eta^a)=G^+(p)+\eta^aF^+_a(p)+\frac{1}{2}\eta^a\eta^bS_{ab}(p)+\frac{1}{3!}\epsilon_{abcd}\eta^b\eta^c\eta^d
F^a_-(p)+\frac{1}{4!}\epsilon_{abcd}\eta^a\eta^b\eta^c\eta^d G_-(p),
\end{equation}
where $a,b,c,d=1,2,3,4$ and it contains following $2^4= 16$
components: one positive-helicity gluon $G^+$, four
positive-helicity fermions $F^+_a$, six scalars $S_{ab}$ which
satisfy self-duality condition $S^{ab}=\epsilon^{abcd}S_{cd}/2$,
four negative-helicity fermions $F^a_-$, and finally one
negative-helicity gluon $G_-$. Similarly all 256 helicity states in
$\mathcal{N}=8$ SUGRA are unified in a superfield, which depend on
Grassmann variables $\eta^A$ with the $SU(8)$ $R$-symmetry index
$A=1,...,8$.

It is a well-known fact that states of ${\cal N}=8$ theory can be
written as the square of states of ${\cal N}=4$ theory. In other
words, the $SU(8)$ $R$-index $A=1,2,..,8$ can be split into two
$SU(4)$ $R$-index, $\W a=1,2,3,4$ and $a=5,6,7,8$ (see for example
~\cite{Bianchi:2008pu}). This square structure is most transparent
in string theory where closed-string vertex is the the product of
left- and right-hand open-string vertices. Using string theory,
accurate relations of tree-level scattering amplitudes between
gravitons and gluons are given in~\cite{Kawai:1985xq}. The KLT
relations express the superamplitude
$\mathcal{M}_n(\{p_i,\eta^A_i\})$ with total $n!$ symmetry in terms
of product of two color-ordered superamplitudes
$\mathcal{A}_n(\sigma|\{p_i,\eta^a_i\})$ and $\mathcal{\W
A}_n(\sigma|\{p_i,\eta^{\W a}_i\})$. Using these relations, explicit
mapping of states has also been given in~\cite{Bianchi:2008pu}.

As we have mentioned in the introduction, we will not use string
theory to study the relations between gravity theory and Yang-Mill
theory. Instead we will derive and prove their relations in the
frame of S-matrix program. To do this, let us start with the
on-shell three point function~\cite{Nair:1988bq}
\bea {\cal A}_3^{\rm{MHV}} & = & {\delta^{(8)} (\sum_i
\ket{i}\eta_i^a)\over \Spaa{1|2}\Spaa{2|3}\Spaa{3|1}},~~~~~{\cal
A}_3^{\O {\rm{MHV}}} = { \delta^{(4)} (\eta_1^a
\Spbb{2|3}+\eta_2^a\Spbb{3|1}+ \eta_3^a\Spbb{1|2})\over
\Spbb{1|2}\Spbb{2|3}\Spbb{3|1}}~~~\label{SUSY-A3} \eea
for gauge theory and
\bea {\cal M}_3^{\rm{MHV}} & = & {\delta^{(16)} (\sum_i
\ket{i}\eta_i^A)\over (\Spaa{1|2}\Spaa{2|3}\Spaa{3|1})^2},~~~~~{\cal
M}_3^{\O {\rm{MHV}}} = { \delta^{(8)} (\eta_1^A
\Spbb{2|3}+\eta_2^A\Spbb{3|1}+ \eta_3^A\Spbb{1|2})\over
(\Spbb{1|2}\Spbb{2|3}\Spbb{3|1})^2}~~~\label{SUSY-M3} \eea
for gravity. Eq. (\ref{SUSY-A3}) and (\ref{SUSY-M3}) are the
supersymmetric generalizations of Eq. (\ref{A3}) and (\ref{M3}).
However, there is one important difference we want to emphasize.
Supersymmetry does not only group different fields together, it
fixes interactions to some level. The most severe constraints arise
in the ${\cal N}=4$ SYM theory and ${\cal N}=8$ SUGRA theory, where
interactions are completely determined by supersymmetry. Thus
supersymmetry adds the so called "selection rule" for non-vanishing
scattering amplitudes, i.e., they must be $SU({\cal N})$
$R$-symmetry invariant. In other words, comparing to the
non-supersymmetric case, besides the familiar assumptions in our
S-matrix frame, {\sl for supersymmetric case we have added another
assumption, the supersymmetric selection rule}.

Having Eq. (\ref{SUSY-A3}) and (\ref{SUSY-M3}), we can write down
the total amplitude
\bea {\cal A}_3(1,2,3)={\cal A}_3^{\rm{MHV}}(1,2,3)+ {\cal A}_3^{\O
{\rm{MHV}}}(1,2,3)~~~~\label{SUSY-A3-Tot}\eea
for ${\cal N}=4$ SYM theory and
\bea {\cal M}_3(1,2,3)={\cal M}_3^{\rm{MHV}}(1,2,3)+ {\cal M}_3^{\O
{\rm{MHV}}}(1,2,3)~~~~\label{SUSY-M3-Tot}\eea
for ${\cal N}=8$ SUGRA theory. One important result is that
\bea {\cal M}_3(1,2,3)={\cal A}_3(1,2,3){\cal \W
A}_3(1,2,3)~~~\label{KLT-3-point} \eea
where we have used the state mapping in~\cite{Bianchi:2008pu} that
$SU(8)$ index $A=1,...,8$ split into $SU(4)$ index
$\tilde{a}=1,2,3,4$ of $\mathcal{\tilde{A}}_n$ and $SU(4)$ index
$a=5,6,7,8$ of $\mathcal{A}_n$ respectively, as well as the
supersymmetric generalization of (\ref{A3-vanish})
\bea {\cal A}_3^{MHV}(1,2,3) {\cal A}_3^{\O
{MHV}}(1,2,3)=0~.~~~\label{SUSU-A3-vanish}\eea
It is worth to emphasize that Eq. (\ref{KLT-3-point}) unifies both
KLT relations (\ref{M3}) as well as the vanishing identity
(\ref{A3-vanish}) for three-point amplitudes in the S-matrix frame.
Then using the BCFW on-shell recursion relations we will generalize
this result to general $n$, thus unify the results presented in
~\cite{BjerrumBohr:2010ta,BjerrumBohr:2010zb}. Since most details
can be found in~\cite{BjerrumBohr:2010ta,BjerrumBohr:2010zb}, our
discussion will be brief.

\subsection{Super-KLT relations with manifest $(n-2)!$ permutation symmetries}

The super-KLT relations with  manifest $(n-2)!$ permutation
symmetries for $\mathcal{N}=8$ SUGRA and $\mathcal{N}=4$ SYM can be
written  as following
\begin{equation}\label{superKLT}
\mathcal{M}_n(\{p_i,\eta^A_i\})=\frac{1}{s_{12...(n-1)}}\sum_{\gamma,\beta\in
S_{n-2}}\tilde{\mathcal{A}}_n(n,\gamma,1|\{p_i,\eta^{\tilde{a}}_i\})
{\cal S}[\gamma|\b]_{p_1}\mathcal{A}_n(1,\beta,n|\{p_i,\eta^a_i\}).
\end{equation}
where again we have split $SU(8)$ index $A=1,...,8$  into $SU(4)$
index $\tilde{a}=1,2,3,4$ of $\mathcal{\tilde{A}}_n$ and $SU(4)$
index $a=5,6,7,8$ of $\mathcal{A}_n$ respectively.
The kinematic
invariants are defined as $s_{K}=(\sum_{i\in K} p_i)^2$ for any
index set $K\subseteq \{1,...,n\}$, and the functional ${\cal S}$ is
defined as ~\cite{BjerrumBohr:2010ta,BjerrumBohr:2010zb}
\begin{equation}
{\cal
S}[i_1,...,i_m|j_1,...,j_m]_{p_1}=\prod^m_{t=1}(s_{i_t1}+\sum^m_{q>t}\theta(i_t,i_q)s_{i_ti_q}),
\end{equation}
where $\theta(i_a,i_b)$ is $1$ if $i_a$ appears after $i_b$ in the
sequence $\{j_1,...,j_m\}$, otherwise it is $0$. The functional
${\cal S}$ has some nice properties. For example,
\begin{equation}
{\cal S}[i_1,...,i_m|j_1,...,j_m]={\cal S}[j_m,...,j_1|i_m,...,i_1],
\end{equation}
which ensures that Eq.~(\ref{superKLT}) is completely symmetric in
${\cal A}_n$ and ${\cal \W A}_n$. More importantly, ${\cal S}$ has
the factorization,
\begin{equation}
{\cal S}[\gamma \sigma|\alpha \beta]={\cal S}[\sigma|\alpha]{\cal
S}_{P}[\gamma|\beta],
\end{equation}
where $P=\sum_{i\in \{\sigma\}}p_i$ has been put on-shell, i.e.,
$P^2=0$.

Due to the appearance of on-shell singularity from
$s_{12...(n-1)}=p_n^2$, Eq.~(\ref{superKLT}) is well-defined only
after regularization. The details of the regularization and concrete
examples of these relations, can be found
in~\cite{BjerrumBohr:2010ta,KLT_long}.

Formula (\ref{superKLT}) is different from the well-known KLT
formula presented in~\cite{Kawai:1985xq,Bern:1998ug} with only
manifest $(n-3)!$ permutation symmetry. As will be shown in
~\cite{KLT_long}, there exist a family of more compact KLT relations
with manifest $(n-3)!$ symmetric form, which do not need any
regularization. The original ansatz in~\cite{Bern:1998ug} is a
special case of these relations, and since we do not bother to write
down the most general form, we will present only following more
symmetric form
%
\begin{equation}\label{superKLTmin}
\mathcal{M}_n(\{p_i,\eta^A_i\})=\sum_{\gamma,\beta\in
S_{n-3}}\tilde{\mathcal{A}}_n(n-1,n,\gamma,1|\{p_i,\eta^{\tilde{a}}_i\})
{\cal
S}[\gamma|\b]_{p_1}\mathcal{A}_n(1,\beta,n-1,n|\{p_i,\eta^a_i\}),
\end{equation}
where $\gamma$ and $\beta$ are permutations of legs $2,...,n-2$.
Note that both forms of KLT relations for superamplitudes are the
same as those for pure gluon and graviton amplitudes.

As will be discussed in detail in~\cite{KLT_long}, it is not easy to
prove (\ref{superKLTmin}) without assumption of total symmetric
property, i.e., the formula (\ref{superKLTmin}) is in fact $n!$
permutation symmetric (or at least $(n-2)!$ permutation symmetric),
but the relation (\ref{superKLT}) is much easier to prove using
supersymmetric BCFW recursion relations
~\cite{ArkaniHamed:2008gz,Brandhuber:2008pf} in $\mathcal{N}=8$
SUGRA and $\mathcal{N}=4$ SYM(we use $\mathcal{M}_n$ to denote
superamplitudes in both theories)
\begin{equation}\label{superBCFW}
\mathcal{M}_n=\sum_{L,R}\int d^{\mathcal{N}}\eta
\mathcal{M}_L(\hat{1},...,\{-\hat{P},\eta\})\frac{1}{P^2}
\mathcal{M}_R(\{\hat{P},\eta\},...,\hat{n}),
\end{equation}
where as in~\cite{BjerrumBohr:2010ta}, we have picked legs $1$ and
$n$ to deform
\begin{equation}
\lambda_1(z)=\lambda_1+z\lambda_n,~~
\tilde{\lambda}_n(z)=\tilde{\lambda}_n-z\tilde{\lambda}_1,~~
\eta_n(z)=\eta_n-z\eta_1.
\end{equation}
It is worth to remind that for supersymmetric case
~\cite{ArkaniHamed:2008gz}, the choice of deformation does not
depend on the helicity of $1$ and $n$.

As expected, the proof follows exactly the same steps as that for
pure graviton and gluon
amplitudes~\cite{BjerrumBohr:2010ta,BjerrumBohr:2010zb,KLT_long}.
The only difference is that we need to replace all helicity sums for
graviton and gluon amplitudes by Grassmann integrations for
superfields.

Now we prove both forms (\ref{superKLT}) and (\ref{superKLTmin}) of
KLT relations by induction. The case $n=3$ has been carefully
discussed in previous paragraphs using only some general properties
of supersymmetric field theory. Assuming both forms hold for any
lower-point superamplitudes up to $n-1$ point, for the $n$-point
case, we uses Eq.~(\ref{superBCFW}) to expand both superamplitudes
$\tilde{A}_n$ and $A_n$, on the R.H.S. of Eq.~(\ref{superKLT}) and
Eq.~(\ref{superKLTmin}), in terms of lower-point superamplitudes. As
discussed in~\cite{BjerrumBohr:2010ta,KLT_long} there are two
classes of contributions,
\begin{itemize}
\item The pole appears in only one of the amplitudes $\tilde{\mathcal{A}}_n$
and $\mathcal{A}_n$.
\item The pole appears in both amplitudes $\tilde{\mathcal{A}}_n$ and $\mathcal{A}_n$.
\end{itemize}

Similar to the discussions in~\cite{BjerrumBohr:2010ta}, for
Eq.~(\ref{superKLT}), when only $\mathcal{A}_n$ has the pole, we
obtain
\begin{eqnarray}
&&\frac{(-)^n}{s_{\hat{1}2...(n-1)}}\sum_{\gamma,\sigma,\beta}\frac{\int
d^4\eta
\tilde{\mathcal{A}}_{n-k+1}(\hat{n},\gamma,\{-\hat{P},\eta\})
\mathcal{\tilde{A}}_k(\{\hat{P},\eta\},\sigma,\hat{1})}{s_{12...k}}
{\cal
S}[\gamma\sigma|\beta]\mathcal{A}_n(\hat{1},\beta,\hat{n})\\\nonumber
&&\propto
\sum_{\sigma}\tilde{\mathcal{A}}_{k+1}(\{\eta,\hat{P}\},\sigma,\hat{1})S[\sigma|\rho]=0,
\end{eqnarray}
where we have used the factorization property of ${\cal S}$, and
$\rho$ is the relative ordering of legs $2,...,k$ in $\beta$.
Similarly the contribution when only $\tilde{\mathcal{A}}_n$ having
the pole also vanishes.

When there are poles in both amplitudes of Eq.~(\ref{superKLT}), the
result is,
\begin{eqnarray}
&&\frac{(-)^n}{s_{\hat{1}2...(n-1)}}\sum_{\gamma,\beta,\sigma,\alpha}\frac{\int
d^4\eta
\tilde{\mathcal{A}}(\hat{n},\gamma,\{-\hat{P},\eta\})\tilde{\mathcal{A}}(\{\hat{P},\eta\},\sigma,\hat{1})}
{s_{12...k}}S[\gamma\sigma|\alpha\beta]\frac{\int d^4\eta
\mathcal{A}(\hat{1},\alpha,\{-\hat{P},\eta\})\mathcal{A}(\{\hat{P},\eta\},\beta,\hat{n})}{s_{\hat{1}2...k}}
\nonumber \\\nonumber &&=\frac{(-)^n}{s_{12...k}}\int d^8\eta
\sum_{\sigma,\alpha}\frac{\tilde{\mathcal{A}}(\{-\hat{P},\eta\},\sigma,\hat{1})
{\cal S}[\sigma|\alpha]\mathcal{A}(\hat{1},\alpha,\{\hat{P},\eta\})}
{s_{\hat{1}2...k}}\sum_{\gamma,\beta}\frac{\tilde{\mathcal{A}}(\hat{n},\gamma,\{-\hat{P},\eta\})
{\cal
S}_{\hat{P}}[\gamma|\beta]\mathcal{A}(\{\hat{P},\eta\},\beta,\hat{n})}{s_{\hat{P}k+1...(n-1)}}\\\nonumber
&&=\frac{\int d^8\eta
\mathcal{M}_{k+1}(\hat{1},...k,\{-\hat{P},\eta\})\mathcal{M}_{n-k+1}
(\{\hat{P},\eta\},k+1,...,\hat{n})}{s_{1...k}},~~~\label{SUSY-n-2-proof}
\end{eqnarray}
where in the first equality we have used
$s_{\hat{1}2...k}=s_{\hat{P}k+1...(n-1)}$ and the factorization
property of ${\cal S}$, and we have combined two $\mathcal{N}=4$
Grassmann integrations over $\eta^{\tilde{a}}$ and $\eta^a$ into a
single $\mathcal{N}=8$ integration over $\eta^A$; while in the
second equality we have used the KLT relations for lower-point
superamplitudes.

Therefore, by Eq.~(\ref{superBCFW}) for $\mathcal{N}=8$ SUGRA, we
have proved the validity of $n$-point KLT relations,
Eq.~(\ref{superKLT}). For manifestly $(n-3)!$ permutation symmetric
KLT relations like Eq.~(\ref{superKLTmin}), as will be discussed
in~\cite{KLT_long}, one has to consider different pole structure,
i.e., if we take $1,n$ to do the BCFW-deformation, then the pole
includes both $p_1,p_{n-1}$ will be very hard to prove. For this
kind of difficult pole structures, the total $n!$ permutation
symmetry (or at least the $(n-2)!$ permutation symmetry) has been
assumed to avoid the direct proof of this kind of poles. Although
this assumption is right from the point of view of string theory, as
far as we know, there is no direct proof in field theory and it will
be very interesting to do that, using, for example, the BCJ
relations. Under this assumption, by simply using Grassmanian
integrations instead of helicity sums, a similar proof for this form
(\ref{superKLTmin}) of KLT relations is straightforward and we shall
not repeat it here.

Note that in the case of pure gluon and graviton
amplitudes~\cite{BjerrumBohr:2010ta,KLT_long}, there are the
so-called mixed-helicity terms, which have to vanish for the use of
usual BCFW recursion relations. In~\cite{BjerrumBohr:2010zb}, new
gluon amplitude relations were found which include the vanishing
results for mixed-helicity terms. In our proof using
superamplitudes, these mixed-helicity terms are unified with normal
terms in the supersymmetric BCFW recursion relations as we have
demonstrated for the three-point superamplitude. As a result, we
shall see immediately that Eq.~(\ref{superKLT}) and
Eq.~(\ref{superKLTmin}) encode not only KLT relations in
maximally-supersymmetric theories, but also new relations among
amplitudes in $\mathcal{N}=4$ SYM, which are generalizations of the
gluon amplitude identities in~\cite{BjerrumBohr:2010zb}.

\section{New relations among amplitudes in $\mathcal{N}=4$ SYM}

As mentioned before, the SYM superamplitudes $\tilde{\mathcal{A}}_n$
and $\mathcal{A}_n$ can be expanded as polynomials of
$\eta^{\tilde{a}}_i$ and $\eta^a_i$ ($i=1,...,n$) respectively, and
the coefficients are component amplitudes with all possible external
helicity states, or matter contents,
\begin{equation}
\mathcal{A}_n=\sum_{\{I_i\}}\prod^{n}_{i=1}\eta^{I_i}_i
A_n(\{I_i\}),
\end{equation}
where $I_i\subseteq\{5,6,7,8\}$ are 16 possible powers of
$\eta^a_i$, which are in one-to-one correspondence with 16 possible
external states in Eq.~(\ref{multiplet}). For example, if
$I_i=\{\{5,6,7,8\},\{5,6,7,8\},\emptyset,\emptyset\}$, then
$A_4(\{I_i\})=A(\{G_-,G_-,G^+,G^+\})$; for
$I_i=\{\{5,6\},\emptyset,\{7,8\},\{5,6,7\},\emptyset,\{8\}\}$, we
have $A_6(\{I_i\})=A(\{S_{56},G^+,S_{78},F_-^8,G^+,F^+_8\})$.
Similarly, for $\tilde{\mathcal{A}}_n$, we have,
\begin{equation}
\tilde{\mathcal{A}}_n=\sum_{\{\tilde{I}_i\}}\prod^{n}_{i=1}\eta^{\tilde{I}_i}_i
A_n(\{\tilde{I}_i\}),
\end{equation}
where $\tilde{I}_i\subseteq\{1,2,3,4\}$.

Plugging both expansions into Eq.~(\ref{superKLT}) and
(\ref{superKLTmin}), it is straightforward to see that the L.H.S.
can be expanded as a polynomial of $\eta^A_i$,
\begin{equation}
\mathcal{M}_n=\sum_{\{J_i\}}\prod^{n}_{i=1}\eta_i^{J_i}M_n(\{J_i\}),
\end{equation}
where $J_i=\tilde{I}_i\cup I_i\subseteq\{1,...,8\}$ are 256 possible
powers of $\eta^A_i$, corresponding to 256 external states in the
$\mathcal{N}=8$ multiplet. The component amplitudes $M_n(\{J_i\})$
satisfy the component KLT relations which directly follow from
Eq.~(\ref{superKLT})
\begin{equation}\label{compKLT}
M_n(\{J_i=\tilde{I}_i\cup
I_i\})=\frac{1}{s_{12...(n-1)}}\sum_{\gamma,\beta\in
S_{n-2}}\tilde{A}_n(n,\gamma,1|\{\tilde{I}_i\}){\cal
S}[\beta|\gamma]A_n(1,\beta,n|\{I_i\}),
\end{equation}
or  from Eq.~(\ref{superKLTmin})
\begin{equation}\label{compKLTmin}
M_n(\{J_i=\tilde{I}_i\cup I_i\})=\sum_{\gamma,\beta\in
S_{n-3}}\tilde{A}_n(n-1,n,\gamma,1|\{\tilde{I}_i\}){\cal
S}[\beta|\gamma]A_n(1,\beta,n-1,n|\{I_i\}).
\end{equation}

However, there are more relations in Eq.~(\ref{compKLT}) and
Eq.~(\ref{compKLTmin}) than the usual gravity KLT relations and
their matter
generalizations~\cite{Kawai:1985xq,Bern:1998ug,Bern:1999bx}, because
$M_n(\{J_i\})$ can vanish even when both $\tilde{A}_n$ and $A_n$ are
non-zero amplitudes. In this case, Eq.~(\ref{compKLT}) and
Eq.~(\ref{compKLTmin}) represent new relations among amplitudes in
$\mathcal{N}=4$ SYM.

We now determine the sufficient and necessary conditions for the
appearance of such relations. The key point is that the
superamplitude must be invariant under the $SU({\cal N})$
$R$-symmetry, as we have mentioned before. In $\mathcal{N}=4$ SYM,
$SU(4)$ $R$-symmetry puts a constraint on any coefficient of the
$\eta$ expansion: it is non-zero if and only if there are same
numbers of $\eta^1,\eta^2,\eta^3$ and $\eta^4$ in that term and each
of them appear at least $2$ times and at most $n-2$ times (the
three-point amplitude is an exception). In other words, if we define
the number of $1,2,3$ and $4$ in $\{\tilde{I}_i\}$ as $n_1,n_2,n_3$
and $n_4$ respectively, then the component amplitude
$\tilde{A}_n(\{\tilde{I}_i\})$ is non-zero if and only if $2\leq
n_1=n_2=n_3=n_4=\W k\leq n-2$. We denote this number by $\tilde{k}$,
it is straightforward to see that $4\tilde{k}$ is the total degree
of $\eta$ and $n-2\tilde{k}$ has the interpretation as the sum of
helicities of $n$ external states.  Similarly $A_n(\{I_i\})$ is
non-zero if and only if $2\leq n_5=n_6=n_7=n_8\leq n-2$, and we
denote the number by $k$.

It is well known that the number $k$ represents a component
amplitude belonging to the N${}^{k-2}$MHV sector, and schematically
we have an expansion in terms of $k$ for the superamplitudes
$\mathcal{A}_n$,
\begin{equation}\label{kexpansion}
\mathcal{A}_n=\sum^{n-2}_{k=2}A^k_n(\eta)^{4k}=A_n^{\rm{MHV}}(\eta)^8+A_n^{\rm{NMHV}}(\eta)^{12}+...+A_n^{\overline{\rm{MHV}}}(\eta)^{4n-8},
\end{equation}
where we have used the fact that N${}^{n-2}$MHV sector is equivalent
to $\overline{\rm{MHV}}$ sector, and we have similar expansions for
$\tilde{\mathcal{A}}_n$ in terms of $\tilde{k}$.

Since $J_i=\tilde{I}_i\cup I_i$, we know the numbers of $1,2,...8$
are exactly given by $n_1,...,n_8$ in $\tilde{I}_i, I_i$. Now given
$n_1=n_2=n_3=n_4=\tilde{k}$, $n_5=n_6=n_7=n_8=k$ and $2\leq
\tilde{k},k\leq n-2$, we conclude from $SU(8)$ $R$-symmetry that the
necessary and sufficient condition for $M_n(\{J_i\})$ to be non-zero
is $\tilde{k}=k$. In other words, when one uses
Eq.~(\ref{kexpansion}) for SYM superamplitudes in
Eq.~(\ref{superKLT}) and Eq.~(\ref{superKLTmin}), only those
component amplitudes from the same sector $\tilde{k}=k$ give
non-zero result, while all interference terms from $\tilde{k}\neq k$
vanish. Therefore, we have seen that there are non-trivial
identities among amplitudes in $\mathcal{N}=4$ SYM from, say,
Eq.~(\ref{superKLTmin})
\begin{equation}\label{relation}
0=\sum_{\gamma,\beta\in
S_{n-3}}\tilde{A}^{\tilde{k}}_n(n-1,n,\gamma,1)S[\beta|\gamma]A^k_n(1,\beta,n-1,n),
\end{equation}
for any $\tilde{k}\neq k$.

One immediate implication of Eq.~(\ref{relation}) is the identities
for flipped-helicity gluon amplitudes~\cite{BjerrumBohr:2010zb}.
There the number of positive(negative) helicity legs in gluon
amplitude $A_n$ which is changed to negative(positive) helicity legs
in $\tilde{A}_n$ is denoted by $n^+(n^-)$, and it was found that the
L.H.S. of KLT relations using $A_n$ and $\tilde{A}_n$ vanishes if
$n^+\neq n^-$ (one particular example is  the mixed-helicity terms
mentioned before with $(n^+,n^-)=(1,0)$ or $(0,1)$). Now since
$\tilde{k}=k+n^--n^+$ for gluon amplitudes, one can see that new
gauge theory identities found in~\cite{BjerrumBohr:2010zb} are
nicely packed into Eq.~(\ref{relation}).

Of course there are more identities when one includes scalars and
fermions. Whenever we have different values for sums of helicities
in $\tilde{A}_n$ and $A_n$, there is an identity among these
amplitudes. For example,
\begin{equation}
0=\sum_{\gamma,\beta\in
S_3}\tilde{A}_6(5,6,\gamma,1|\{S_{12},G_-,F^+_3,F^+_4,G_-,G^+\})S[\beta|\gamma]A_6(1,\beta,5,6|\{S_{56},G^+,S_{78},F_-^8,G^+,F^+_8\}),
\end{equation}
since the first one is an NMHV amplitude while the second is an MHV
amplitude.

Generally speaking, let us fix the $\W k$ and matter
contents\footnote{Given $\W k$, there are various N${}^{\W k-2}$MHV
amplitudes $\W A_n^{\W k}$ differing from each other by the matter
contents of $n$ particles.} of the N${}^{\W k-2}$MHV amplitude $\W
A^{\W k}_n$, then Eq.~(\ref{relation}) can be considered as
relations among $(n-3)!$ color-ordered N${}^{k-2}$MHV amplitudes
$A^k_n$ with any particular $k\neq \W k$ and matter contents. It is
worth to notice that since the sum in Eq.~(\ref{relation}) is over
$(n-3)!$ basis amplitudes, we obtain a linear relation among these
amplitudes in the basis.

However, the above statement is not in contradiction with the
statement that $(n-3)!$ is the number of minimal basis amplitudes
for gauge theory amplitudes, because for amplitudes in the sector
$k=\tilde{k}$, the same combination gives an amplitude in
$\mathcal{N}=8$ SUGRA which does not vanish. Therefore, unlike
helicity-independent relations such as Kleiss-Kuijf relations and
BCJ relations, these relations only hold for amplitudes in sectors
with $k\neq\tilde{k}$. In other words, new gauge identities tell us
that although the helicity-independent basis consists of $(n-3)!$
amplitudes, for given helicity category, it is possible to reduce
the number of basis amplitudes further. This can be used to speed up
the calculations of cross sections for, for example, the LHC
experiments.

Since for each given  N${}^{\W k-2}$MHV amplitude $\W A_n^{\W k}$
with particular matter contents, Eq.~(\ref{relation}) gives a linear
relation among $(n-3)!$ color-ordered N ${}^{k-2}$MHV amplitudes
$A_n^k$, thus with different choices of $\W k$ and matter contents
for $\W A_n$, we get different linear equations for same basis set
$A_n^{N^{k-2}MHV}$. In other words, for any fixed $k$ and matter
contents, Eq.~(\ref{relation}) represents a huge number of relations
among these specific $(n-3)!$ amplitudes. Thus it would be very
interesting to see how powerful the constraints from
Eq.~(\ref{relation}) in reducing the number of independent
color-ordered amplitudes\footnote{Given any $\tilde{k}\neq k$,
amplitudes with different external states are related by
supersymmetric Ward identities. A basis for component amplitudes has
been found in~\cite{Elvang:2009wd}, from which $A^{\tilde{k}}_n$
with any matter contents can be obtained. We conjecture that for any
$2\leq\tilde{k}\leq n-2$ and $\tilde{k}\neq k$, there is one
independent relation for each basis component amplitude found
in~\cite{Elvang:2009wd}}.

We use a simple example to demonstrate. For $n=5$ and $k=3$, there
is only one non-trivial identity, since one can only choose
$\tilde{k}=2$ and in this MHV case all component amplitudes can be
related to the gluon amplitude
$\tilde{A}_5(\{G_-,G_-,G^+,G^+,G^+\})$. Amplitudes $A^{\rm{NMHV}}_5$
with any external states are also related to the googly amplitude
$A^{\rm{NMHV}}_5(G^+,G^+,G_-,G_-,G_-)$. Thus we have one relation
for the two basis color-ordered amplitudes
$A^{\rm{NMHV}}_5(1,2,3,4,5)$ and $A^{\rm{NMHV}}_5(1,3,2,4,5)$,
\begin{eqnarray}
&&[\frac{s_{21}s_{31}}{\langle45\rangle\langle52\rangle\langle23\rangle\langle31\rangle\langle14\rangle}+\frac{(s_{21}+s_{32})s_{31}}{\langle45\rangle\langle53\rangle\langle32\rangle\langle21\rangle\langle14\rangle}]A^{\rm{NMHV}}(1,2,3,4,5)\\\nonumber
+&&[\frac{(s_{31}+s_{32})s_{21}}{\langle45\rangle\langle52\rangle\langle23\rangle\langle31\rangle\langle14\rangle}+\frac{s_{31}s_{21}}{\langle45\rangle\langle53\rangle\langle32\rangle\langle21\rangle\langle14\rangle}]A^{\rm{NMHV}}(1,3,2,4,5)=0,
\end{eqnarray}
where we have used the expression of MHV amplitude. This relation
reduces the number of independent color-ordered 5-point NMHV
amplitudes to one.

For the special case $\tilde{k}=2$, it is straightforward to derive
these new relations among non-MHV amplitudes explicitly. Since we
have
\begin{equation}
0=\sum_{\gamma,\beta\in
S_{n-3}}\tilde{A}^{\rm{MHV}}_n(n-1,n,\gamma,1)S[\beta|\gamma]A^k_n(1,\beta,n-1,n)
\end{equation}
for any $k>2$.  Plugging the expression of MHV amplitude and using
BCJ relations recursively, we obtain~\cite{FHHJ-paper}
\begin{equation}
0=\sum_{\beta\in
S_{n-3}}\prod^{n-3}_{i=1}[\beta_i|p_{\beta_{i+1}}+p_{\beta_{i+2}}+...+p_{\beta_{n-1}}|n\rangle
A^k_n(1,\beta,n-1,n),
\end{equation}
where each $\beta$ is a permutation of $2,...,n-2$ and $\beta_i$
denotes its $i$-th element.

\section{Conclusion and discussions}

In this short note, we studied the supersymmetric version of KLT
relations, including the $(n-3)!$ symmetric version
~\cite{Kawai:1985xq,Bern:1998ug} and the newly discovered $(n-2)!$
symmetric version~\cite{BjerrumBohr:2010ta,BjerrumBohr:2010zb}, in
the frame of S-matrix program. In this frame, we do not use string
theory or the Lagrangian definition of field theory. Besides the
well-known principles, we have added only following two assumptions:
the validity of BCFW recursion relations and the supersymmetry.

Our main results are two formulae (\ref{superKLT}) and
(\ref{superKLTmin}). The advantage of going to supersymmetric
version is now we unified supersymmetric KLT relations
~\cite{BjerrumBohr:2010ta} and the newly discovered gauge theory
identities~\cite{BjerrumBohr:2010zb} into one frame and produced
more vanishing identities involving the scalars and fermions. As we
have discussed at the end of previous section, these new identities
imply further reduction of number of helicity basis of various
amplitudes. One obvious project is to study how many linearly
independent relations we can obtain from these new identities.

The main reason that we can have a unified picture in the frame of
S-matrix program  is  that one extra assumption, i.e., the
supersymmetric selection rule, has been added. As we have emphasized
in previous sections, the added supersymmetry fixed the interactions
among scalars, fermions, gauge bosons and gravitons, thus it is a
very strong extra condition. This is the price we need to pay for
having the unified picture. Amazingly, for tree-level amplitudes of
pure gluons or pure gravitons, results are the same with or without
the supersymmetry. This is why we can lift the theory to the
supersymmetric version and then infer their properties. In other
words, {\sl gluon and graviton know supersymmetry somehow}. For
scalars and fermions, with or without supersymmetry will have huge
differences, so their discussions will be more difficult.

\subsection*{Acknowledgements}

BF is supported by fund from Qiu-Shi, the Fundamental Research Funds
for the Central Universities with contract number 2009QNA3015, as
well as Chinese NSF funding under contract No.10875104. BF, SH
thanks the organizers of the program ``QFT, String Theory and
Mathematical Physics'' at KITPC, Beijing for hospitality while this
work is done.


\begin{thebibliography}{99}


\bibitem{S-matrix} D.I. Olive, Phys. Rev. 135,B 745(1964); G.F.
Chew, "The Analytic S-Matrix: A Basis for Nuclear Democracy",
W.A.Benjamin, Inc., 1966; R.J. Eden, P.V. Landshoff, D.I. Olive,
J.C. Polkinghorne, "The Analytic S-Matrix", Cambridge University
Press, 1966.

\bibitem{Bern:1994zx}
  Z.~Bern, L.~J.~Dixon, D.~C.~Dunbar and D.~A.~Kosower,
  Nucl.\ Phys.\  B {\bf 425}, 217 (1994)
  [arXiv:hep-ph/9403226].

\bibitem{Britto:2004ap}
  R.~Britto, F.~Cachazo and B.~Feng,
  Nucl.\ Phys.\  B {\bf 715}, 499 (2005)
  [arXiv:hep-th/0412308].

\bibitem{Britto:2005fq}
  R.~Britto, F.~Cachazo, B.~Feng and E.~Witten,
  Phys.\ Rev.\ Lett.\  {\bf 94}, 181602 (2005)
  [arXiv:hep-th/0501052].

\bibitem{ArkaniHamed:2008yf}
  N.~Arkani-Hamed and J.~Kaplan,
  JHEP {\bf 0804}, 076 (2008)
  [arXiv:0801.2385 [hep-th]].


\bibitem{boundary}
  B.~Feng, J.~Wang, Y.~Wang and Z.~Zhang,
  JHEP {\bf 1001}, 019 (2010)
  [arXiv:0911.0301 [hep-th]].

  B.~Feng and C.~Y.~Liu,
  arXiv:1004.1282 [hep-th].
\bibitem{Paolo:2007}
  P.~Benincasa and F.~Cachazo,
  arXiv:0705.4305[hep-th].

\bibitem{SHST}
  S.~He and H.~b.~Zhang,
  arXiv:0811.3210 [hep-th];

  P.~C.~Schuster and N.~Toro,
  JHEP {\bf 0906}, 079 (2009)
  [arXiv:0811.3207 [hep-th]].

\bibitem{Feng:2010my}
  B.~Feng, R.~Huang and Y.~Jia,
  arXiv:1004.3417 [hep-th].

\bibitem{Kleiss:1988ne}
  R.~Kleiss and H.~Kuijf,
  Nucl.\ Phys.\  B {\bf 312}, 616 (1989).

\bibitem{Bern:2008qj}
  Z.~Bern, J.~J.~M.~Carrasco and H.~Johansson,
  Phys.\ Rev.\  D {\bf 78}, 085011 (2008)
  [arXiv:0805.3993 [hep-ph]].

\bibitem{BjerrumBohr:2009rd}
  N.~E.~J.~Bjerrum-Bohr, P.~H.~Damgaard and P.~Vanhove,
  Phys.\ Rev.\ Lett.\  {\bf 103}, 161602 (2009)
  [arXiv:0907.1425 [hep-th]].


\bibitem{Stieberger:2009hq}
  S.~Stieberger,
  arXiv:0907.2211 [hep-th].

\bibitem{Sondergaard:2009za}
T.~Sondergaard,
0903.5453 [hep-th].

\bibitem{Jia:2010nz}
  Y.~Jia, R.~Huang and C.~Y.~Liu,
  arXiv:1005.1821 [hep-th].

\bibitem{Benincasa:2007qj}
  P.~Benincasa, C.~Boucher-Veronneau and F.~Cachazo,
  JHEP {\bf 0711}, 057 (2007)
  [arXiv:hep-th/0702032].


\bibitem{Kawai:1985xq}
  H.~Kawai, D.~C.~Lewellen and S.~H.~H.~Tye,
  Nucl.\ Phys.\  B {\bf 269}, 1 (1986).


\bibitem{BjerrumBohr:2010ta}
  N.~E.~J.~Bjerrum-Bohr, P.~H.~Damgaard, B.~Feng and T.~Sondergaard,
  arXiv:1005.4367 [hep-th].



\bibitem{BjerrumBohr:2010zb}
  N.~E.~J.~Bjerrum-Bohr, P.~H.~Damgaard, B.~Feng and T.~Sondergaard,
  arXiv:1006.3214 [hep-th].

\bibitem{Bianchi:2008pu}
  M.~Bianchi, H.~Elvang and D.~Z.~Freedman,
  JHEP {\bf 0809}, 063 (2008)
  [arXiv:0805.0757 [hep-th]].


\bibitem{ArkaniHamed:2008gz}
  N.~Arkani-Hamed, F.~Cachazo and J.~Kaplan,
  arXiv:0808.1446 [hep-th].


\bibitem{Drummond}
 J.M.Drummond, J.Henn, G.P.Korchemsky and E.Sokatchev,
  Nucl.\ Phys.\  B {\bf 828}(2010)317
  [arXiv:0807.1095 [hep-th]].


\bibitem{Nair:1988bq}
  V.~P.~Nair,
  Phys.\ Lett.\  B {\bf 214}, 215 (1988).

\bibitem{Bern:1998ug}
  Z.~Bern, L.~J.~Dixon, D.~C.~Dunbar, M.~Perelstein and J.~S.~Rozowsky,
  Nucl.\ Phys.\  B {\bf 530}, 401 (1998)
  [arXiv:hep-th/9802162].


\bibitem{KLT_long}
  N.~E.~J.~Bjerrum-Bohr, P.~H.~Damgaard, B.~Feng and T.~Sondergaard,
  to appear soon.




\bibitem{Brandhuber:2008pf}
  A.~Brandhuber, P.~Heslop and G.~Travaglini,
  Phys.\ Rev.\  D {\bf 78}, 125005 (2008)
  [arXiv:0807.4097 [hep-th]].

\bibitem{Bern:1999bx}
  Z.~Bern, A.~De Freitas and H.~L.~Wong,
  Phys.\ Rev.\ Lett.\  {\bf 84}, 3531 (2000)
  [hep-th/9912033].
  N.~E.~J.~Bjerrum-Bohr and O.~T.~Engelund,
  Phys.\ Rev.\  D {\bf 81}, 105009 (2010)
  [1002.2279 [hep-th]].

\bibitem{Elvang:2009wd}
  H.~Elvang, D.~Z.~Freedman and M.~Kiermaier,
  arXiv:0911.3169 [hep-th].

\bibitem{FHHJ-paper}
    Bo Feng, Song He, Rijun Huang, Yin Jia, to appear soon.


\end{thebibliography}
\end{document}